\documentclass[doublecol]{epl2}

\title{Enhanced Rashba effect for hole states in a quantum dot}

\author{Aram Manaselyan and Tapash Chakraborty }
\shortauthor{A. Manaselyan, T. Chakraborty}

\institute{ Department of Physics and Astronomy, University of
Manitoba, Winnipeg, Canada R3T 2N2 }

\pacs{73.21.La}{Quantum dots} \pacs{71.70.Ej}{Spin-orbit coupling}
\pacs{75.75.+a}{Magnetic properties of nanostructures}

\abstract{ The effect of Rashba spin-orbit (SO) interaction on the
hole states in a quantum dot is studied in the presence of an
external magnetic field. We demonstrate here that the Rashba SO
coupling has a profound effect on the energy spectrum of the holes
revealing level repulsions between the states with the same total
momentum. We also show that the resulting spin-orbit gap is much
larger than the corresponding one for the electron energy levels in
a quantum dot. Inter-hole interactions only marginally reduce the
spin-orbit gap. This enhanced Rashba effect would manifest itself in
the tuneling current which depends on the spin-orbit coupling
strength. }

\begin{document}

\maketitle

\section{Introduction}

Semiconductor quantum dots are the nanoscale zero-dimensional
systems with discrete energy levels, much like in atoms (and hence
the popular name, {\it artificial atoms} \cite{qdbook}). They have
one great advantage that their shape and the number of electrons in
those systems can be controlled externally and as a result, they
have been the subject of intense research in recent years. They are
particularly promising as components of futuristic devices for
quantum information processing \cite{quantum_info} and for coherent
spin transport \cite{spintro}. The spin states of these systems are
ideal for applications because of their relative insensitivity to
electrical noise in a device environment \cite{bandy}. One proposed
mechanism for coherent spin manipulation in quantum nanostructures
is via the Rashba spin-orbit (SO) coupling \cite{Rashba,book}. The
SO interaction can arise in a quantum dot due to confinement and
lack of inversion symmetry of the nanostructure which creates a
local electric field perpendicular to the electron plane
\cite{Falko, Governale}. The SO coupling strength can be varied by
changing the asymmetry of the quantum structure with an external
electric field. The magnetic field effects on the properties of
low-dimensional systems, such as quantum wells and quantum dots with
the Rashba interaction has been reported in experiments \cite{nitta}
and theory \cite{Ulloa}. In our work on Rashba effects in electron
dots \cite{chakraborty}, we found multiple level crossings and level
repulsions that resulted from the interplay between the Zeeman and
the SO couplings. Level anticrossings observed in quantum
nanostructures have been attributed to the presence of SO coupling
in those systems \cite{level_cross}. However, studies of the Rashba
effect on quantum dots as yet, are limited only to the case of
electrons as charge carriers.

The importance of holes in semiconductor spintronics is well
documented \cite{hole_spintro} in the literature. Some theoretical
results about hole states in quantum dots have been reported earlier
\cite{Rossler, Broido, Pedersen2}. Experiments on hole levels in
quantum dots have also been reported recently
\cite{Maan1,Maan2,others}. Rashba effect is expected to be stronger
in $p$-type quantum wells \cite{ekenberg_1}. Interestingly, for
holes the Dresselhaus effect (due to bulk inversion asymmetry) is
small compared to the Rashba effect \cite{ekenberg_2}. In this
Letter, we report on our studies involving the hole levels in planar
quantum dots with Rashba SO interactions. We find that the SO gap at
the anticrossings of the energy levels is much larger than those for
the case of electrons. Inter-particle interactions (Coulomb type)
reduce the gap somewhat, but it is still orders of magnitude larger
than the corresponding ones for electrons.

The Rashba effect for holes is different from that for
the electrons. It is well known that the electron Rashba
coefficient increases nearly linearly with an increase
of the electric field. But the two-dimensional heavy hole
systems in single heterostructures exhibit a decrease of
Rashba SO splitting with an increase of the electric
field \cite{Winkler}. The effect for a two-dimensional
light hole system is however the same as that for the
electrons. Determination of the Rashba coefficient for
holes was reported for the InP quantum wires \cite{Zhang}
where the band mixing was taken into account. The result
was that as the electric field increases, the hole Rashba
coefficient increases at first, then decreases. There
is also a critical electric field for which the hole
Rashba coefficient vanishes. No such calculations have
been reported as yet, for the hole Rashba coefficients
in quantum dots. However electric field still remains an
useful tool to control the Rashba coefficient.

\section{Theory}

Let us consider the hole states in a InAs/GaAs
cylindrical quantum dot in the presence of an external
magnetic field directed along the $z$ axis. Taking
into account only the $\Gamma_8$ states which correspond
to the states with hole spin $J=3/2$, we can construct
the single-hole Hamiltonian of the system as
\begin{equation}\label{Ham}
{\cal H}={\cal H}_L+{\cal H}_Z+V_{\rm conf}(\rho,z)+
{\cal H}_{\rm SO}.
\end{equation}
Here ${\cal H}_L$ is the Luttinger hamiltonian in axial
representation obtained with the four-band \textbf{k$\cdot$p}
theory \cite{Lutt2,Pedersen2}
\begin{equation}\label{HLut}
{\cal H}_L=\frac{1}{2m_0}\left( \begin{array}{cccc}
{\cal H}_h & R & S & 0 \\
R^* & {\cal H}_l & 0 & S \\
S^* & 0 & {\cal H}_l & -R \\
0 & S^* & -R^* & {\cal H}_h
\end{array}\right),
\end{equation}
where
\begin{eqnarray*}
{\cal H}_h & = & (\gamma_1+\gamma_2)(\Pi_x^2+\Pi_y^2)+
(\gamma_1-2\gamma_2)\Pi_z^2,\\
{\cal H}_l & = & (\gamma_1-\gamma_2)(\Pi_x^2+\Pi_y^2)+
(\gamma_1+2\gamma_2)\Pi_z^2,
\end{eqnarray*}
$R=2\sqrt3\gamma_3{\rm i}\Pi_-\Pi_z, \quad S=\sqrt3\gamma
\Pi_-^2, \quad \gamma=\frac12(\gamma_2+\gamma_3),$
and ${\bf \Pi}={\bf p}-\frac{e}{c}{\bf A}, \quad
\Pi_\pm=\Pi_x\pm{\rm i}\Pi_y.$
$\gamma_1, \gamma_2$ and $\gamma_3$ are the Luttinger
parameters and $m_0$ is the free electron mass. The
Hamiltonian is presented in the hole picture, where the
energies are positive, ${\bf A}$ is the usual symmetric
gauge vector potential. The Zeeman Hamiltonian ${\cal H}_Z$
is a $4\times4$ diagonal matrix with diagonal elements
$-\kappa \mu_B B j_z$, where $\kappa$ is the fourth
Luttinger parameter, $\mu_B$ is the Bohr magneton and
$j_z$ is the projection of the hole spin on the $z$ axis
($j_z=3/2,1/2,-1/2,-3/2$).

\begin{figure}
\onefigure{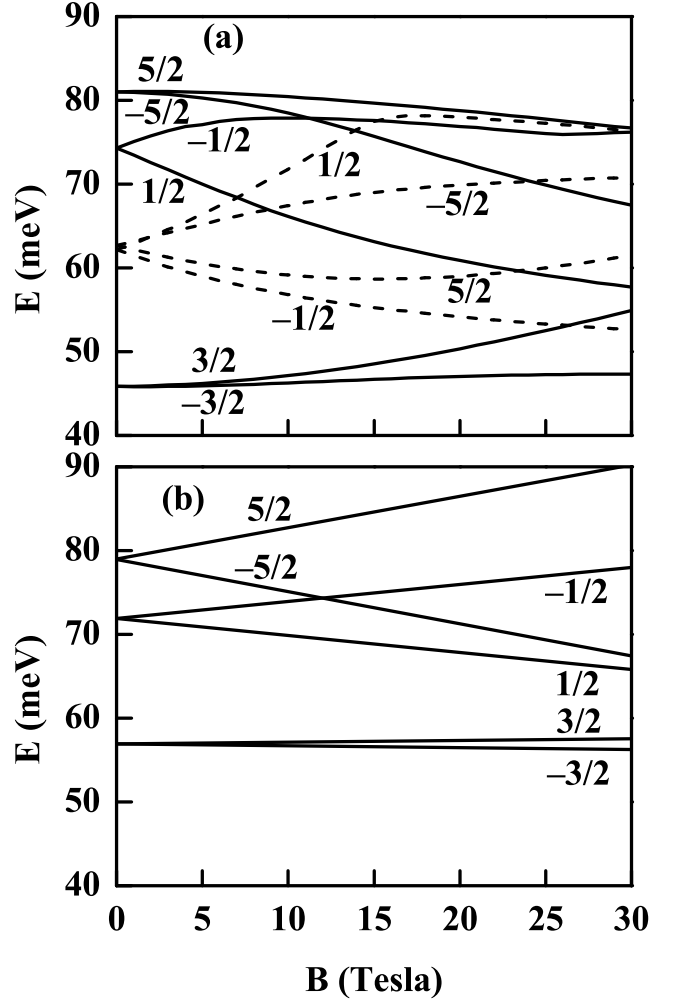} \caption{Magnetic field dependence of single
hole energy levels in the absence of spin-orbit coupling. (a) Our
theoretical results: Solid lines have even parity and dashed lines
have odd parity. (b) Results derived from the experimental data
\cite{Maan2}. The curves are labelled by their corresponding total
momentum $F_z$.} \label{Maan}
\end{figure}

We choose the lateral confinement potential of the dot
as parabolic with oscillator energy $\hbar\omega_0$.
This can be justified from the energies of far-infrared
absorption on such dots which show only a weak dependence
on the electron occupation \cite{Parabol}. We also
take into account the confinement potential in the
$z$ direction as a rectangular well of width $L$.
Just as for electron states in parabolic quantum
dots \cite{chakraborty}, the term ${\cal H}_{\rm SO}$ describes
the Rashba spin-orbit coupling \cite{Rashba} due to the
inhomogeneous potential that confines the holes in a plane
\begin{equation}\label{HSO}
{\cal H}_{\rm SO}=\frac{\alpha}{\hbar}[{\bf J} \cdot {\bf
\Pi}]_z=\frac{\alpha}{\hbar} (J_x\Pi_y-J_y\Pi_x),
\end{equation}
where $\alpha$ is the Rashba coefficient for the holes and
$J_x,J_y$ are the hole spin matrices
\begin{eqnarray*}
J_x&=&\left(\begin{array}{cccc} 0 & {\rm i}\sqrt{3}
& 0 & 0\\
-{\rm i}\sqrt3 & 0 & 2i & 0\\ 0 & -2i & 0 & {\rm i}\sqrt3\\
0 & 0 & -{\rm i}\sqrt3 & 0 \\
\end{array}\right), \\
J_y&=&\left(\begin{array}{cccc} 0 & \sqrt{3} &
0 & 0\\
\sqrt3 & 0 & 2 & 0\\ 0 & 2 & 0 & \sqrt3\\ 0 &
0 & \sqrt3 & 0 \\
\end{array}\right).
\end{eqnarray*}
It is well known \cite{WinklerBook} that in two-dimensional systems
for heavy holes the Rashba interaction is cubic in momentum because
the linear term is zero and we have to keep the higher order terms.
In the case of quantum wires and quantum dots it is not possible to
separate the heavy and light hole states any more due to strong band
mixing effects. Then the linear term in Rashba interaction will not
be zero and we can neglect the cubic terms which are much smaller.
The final form of the spin-orbit coupling hamiltonian is then
$${\cal H}_{\rm SO}=-\frac{\alpha}{\hbar}\left(\begin{array}{cccc}0
& \sqrt3\Pi_- & 0 & 0\\ \sqrt3\Pi_+ & 0 & 2\Pi_- & 0\\ 0 & 2\Pi_
+ & 0 & \sqrt3\Pi_- \\ 0 & 0 & \sqrt3\Pi_+ & 0 \end{array}
\right).$$
The Hamiltonian (\ref{Ham}) is rotationally invariant.
Therefore it will be useful to introduce the total
momentum ${\bf F}={\bf J}+{\bf L}$, where ${\bf J}$
is the angular momentum of the band edge Bloch function,
and ${\bf L}$ is the envelop angular momentum. Since
the projection of the total momentum $F_z$ is a
constant of motion, we can find simultaneous eigenstates
for (\ref{Ham}) and $F_z$ \cite{Sersel}. Therefore for
a given value of $F_z$ a general hole state can be
written as \cite{Pedersen2}
$$\Psi_{F_z}(\rho,\theta,z)=\sum_{j_z}\psi_{j_z}(\rho,z)
{\rm e}^{{\rm i}(F_z-j_z)\theta}|3/2,j_z\rangle,$$
where $|3/2,j_z\rangle$ is the Bloch function,
$\psi_{j_z}(\rho,z)$ is the envelop function and instead
of the angular momentum quantum number $l=0,\pm1,\ldots$,
we write its allowed values $l=F_z-j_z$.

First, we solve the problem only for ${\cal H}_h$
which corresponds to the state $|3/2,3/2\rangle$. For
the in-plane problem we get the equation of a two-dimensional
harmonic oscillator in a magnetic field, the solution of
which is
\begin{equation}\label{WFro}
f_{nl}(\rho,\theta)=C_{nl}\left(\frac{{\rm i}\rho}a
\right)^{|l|} {\rm e}^{-\rho^2/2a^2}L_n^{|l|}\left(
\frac{\rho^2}{a^2}\right){\rm e}^{{\rm i}l\theta}.
\end{equation}
In Eq.~(\ref{WFro}) $L_n^l(x)$ is the generalized Laguerre
polynomial,
$C_{nl}=\frac1a\left[\frac{n!}{\pi(n+|l|)!}\right]^{\frac12}, \qquad
a=\left[\frac{\hbar(\gamma_1+\gamma_2)}{m_0\omega}
\right]^{\frac12}$, $\omega=\sqrt{\omega_0^2+\frac14\omega_c^2}$,
$\omega_c=eB(\gamma_1+\gamma_2)/m_0c$ is the cyclotron frequency,
and $n$ is the radial quantum number ($n=0,1,\ldots$). The energy
levels are given by $E_{nl}=2\hbar\omega\left(n+1/2(|l|+1)-l
\omega_c/4\omega\right).$ For the $z$ part of the problem we can use
the results for a one-dimensional rectangular quantum well
\begin{equation}\label{WFz}
g_s(z)=\sqrt{\frac2L}\sin\left[\frac{s\pi}{L}\left(
z+\frac L2\right)\right],
\end{equation}
with energies $E_s=\pi^2\hbar^2s^2(\gamma_1-2 \gamma_2)/2m_0L^2,
\quad s=1,2,3,\ldots$

It is logical to seek the eigenfunctions of Hamiltonian
(\ref{Ham}) as an expansion with the basis functions
(\ref{WFro}) and (\ref{WFz}).
\begin{equation}\label{Bazis}
\Psi_{F_z}(\rho,\theta,z)=\sum_{n,s,j_z}C(n,s,j_z)
f_{n,F_z-j_z}(\rho,\theta)g_s(z)
\end{equation}
or in the corresponding spinor representation.

The matrix elements of the Hamiltonian (\ref{Ham}) can then be
evaluated analytically. All energies and wave functions for the
single hole system are evaluated numerically using the exact
diagonalization scheme \cite{chakraborty}. If we choose $n_{\rm
max}$ number of in-plane basis states and $s_{\rm max}$ number of
basis states for the $z$ direction, we get a matrix of the order
$N=4n_{\rm max}s_{\rm max}$. In that matrix each element of the
Luttinger Hamiltonian has its submatrix of order $n_{\rm max}s_{\rm
max}$. The submatrix of ${\cal H}_h$ will be diagonal, but the
submatrix of ${\cal H}_l$ will not be so because we are using the
eigenfunctions of ${\cal H}_h$ as the basis functions. We have also
considered an interacting two-hole system, where the interaction
matrix elements are formally same as those for electrons
\cite{qdbook,chakraborty}.

\section{Discussion of results}
All computations were carried out for the InAs/GaAs quantum dot with
parameters $\gamma_1=11.01$, $\gamma_2=4.18$, $\gamma_3=4.84$,
$\kappa=1.2$ and the dielectric constant, $\varepsilon=12.4$
\cite{Vurgaftman}. The height of the dot is taken as $L=4.5$ nm and
the in-plane confinement energy $\hbar\omega_0=20$ meV which seems
to be very reasonable \cite{Warburton,Maan2}.

\begin{figure}
\onefigure{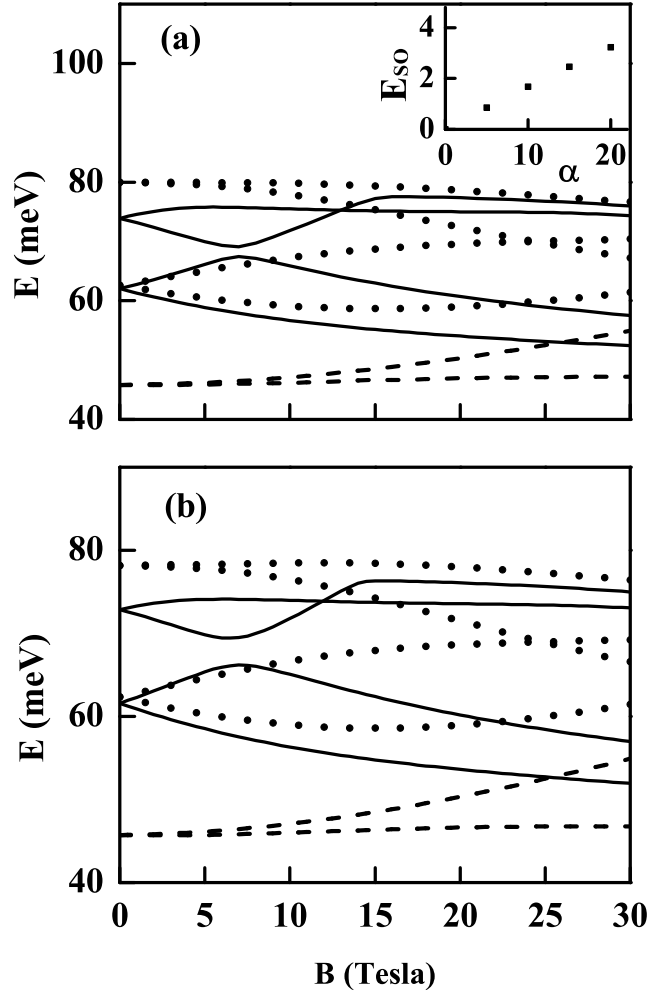} \caption{Dependence of the single hole energy
levels on the magnetic field taking into account the spin-orbit
coupling. (a) $\alpha=10$ meV nm. (b) $\alpha=20$ meV nm. Solid
lines: $F_z=\pm1/2$, dashed lines: $F_z=\pm3/2$, dotted lines:
$F_z=\pm5/2$. The spin-orbit gap $E_{\rm SO}$ vs $\alpha$ is shown
as inset.} \label{alfa}
\end{figure}

First we compare our results for $\alpha=0$ with the results of the
experimental work \cite{Maan2} where the dependence of the single
hole energy levels on the magnetic field was evaluated using the
capacitance-voltage and polarized photoluminescence spectroscopy
measurements in high magnetic fields. In Fig.~\ref{Maan} the
magnetic field dependence of the single hole energy levels are
presented for different values of the total momentum $F_z$. Our
results are shown in Fig.~\ref{Maan}(a), and the ones derived by the
experimental group \cite{Maan2} are shown in Fig.~\ref{Maan}(b). In
both figures the ground state corresponds to $F_z=-3/2$, which lies
below the state with $F_z=3/2$ for all values of the magnetic field.
Clearly, in Fig.~\ref{Maan}(a) there are some levels with
$F_z=\pm1/2$ and $F_z=\pm5/2$ (dashed lines) that are absent in
Fig.~\ref{Maan}(b). The solid lines in Fig.~\ref{Maan}(a) have even
parity while the dashed lines have odd parity \cite{Pedersen2}. We
believe that the absence of the lines in Fig.~\ref{Maan}(b), but
presence in our results is due to the parity selection rule which in
the photoluminescence spectroscopy experiment does not allow the
states presented as dashed lines in Fig.~\ref{Maan}(a). In
Fig.~\ref{Maan}(a) the energies of the states $F_z=\pm5/2$ with even
parity are found to decrease with increasing magnetic field. This is
in contrast to the experimental data [Fig.~\ref{Maan}(b)], where the
energy of the state $F_z=5/2$ increases with increasing magnetic
field. This difference in the magnetic behavior is due to particular
assumptions in \cite{Maan2} that are employed to derive the hole
energy from the experimental data. However, what is crucial here is
the separation between the two curves that is very similar in both
cases. One of the most interesting experimental results of
\cite{Maan2} is the level crossing point around 12 Tesla between the
states $F_z=-1/2$ and $F_z=-5/2$, which is also present in our
results in Fig.~\ref{Maan}(a). Overall, our results for the hole
energy levels (in the absence of Rashba effect) are in quite good
agreement with the results of \cite{Maan2}.

In Fig.~\ref{alfa}, the magnetic field dependence of the single-hole
energy levels is depicted for different values of the total momentum
$F_z$, where the Rashba spin-orbit coupling is taken into account,
with $\alpha=10$ meV nm [Fig.~\ref{alfa}(a)] and $\alpha=20$ meV nm
[Fig.~\ref{alfa}(b)]. In the absence of the spin orbit coupling
(Fig.~\ref{Maan}(a)) the two lowest energy states with total
momentum $F_z=1/2$ cross at a finite magnetic field near 7 Tesla.
Spin orbit interaction mixes the states with same total momentum
$F_z$ and different parities. Hence in Fig.~\ref{alfa}(a), instead
of a level crossing we actually have an `anticrossing' with an
energy gap of $E_{\rm SO}\approx2$ meV; we can no longer separate
the states by parity. With an increase of the spin-orbit coupling
the size of the gap increases and for $\alpha=20$ meV nm
(Fig.~\ref{alfa}(b)) the gap is $E_{\rm SO}\approx4$ meV [see inset
in Fig.~2(a)]. A spin-orbit gap was also found earlier \cite{Hong}
for the electron states in quantum dots and quantum rings, but the
gap was much smaller in comparison to the present case. A comparison
of Fig.~\ref{alfa}(a) and Fig.~\ref{alfa}(b) clearly reveals that
the effect of the spin-orbit coupling is very small for the ground
states with $F_z=\pm3/2$. The reason of this behavior is that the
dominant components of the states $F_z=3/2$ and $F_z=-3/2$ are those
corresponding to $j_z=3/2$ and $j_z=-3/2$ respectively. But the
spin-orbit Hamiltonian (\ref{HSO}) does not mix those states.
Therefore we conclude that the effect of Rashba spin-orbit coupling
is much stronger and more important for the hole states in the
valence band. It mixes states with even and odd parity, removes the
crossing points, and introduces level repulsion between the states
with the same total momemntum $F_z$.

\begin{figure}
\onefigure{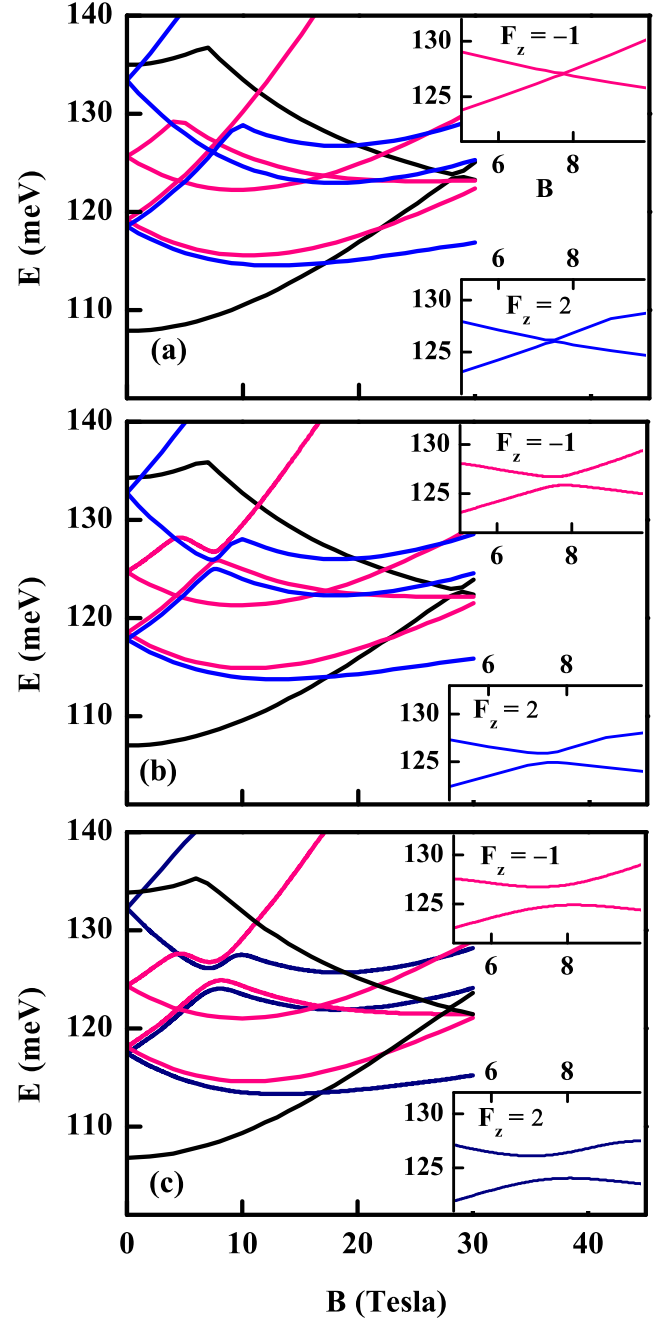} \caption{Dependence of the two holes
(interacting) energy levels on the magnetic field for total momentum
$F_z=0$ (black), $F_z=\pm1$ (pink) and $F_z=\pm2$ (blue). (a)
$\alpha=0$. (b) $\alpha=10$ meV nm. (c) $\alpha=20$ meV nm. Insets
show the level crossing (a) and level repulsion (b,c) gaps for
$F_z=-1$ and $F_z=2$.} \label{coulomb}
\end{figure}

In Fig.~\ref{coulomb} the magnetic field dependence of the two-hole
(interacting) energy levels is depicted for different values of the
total momentum $F_z=0,\pm1,\pm2$. For small values of the magnetic
field, the ground state has total momentum $F_z=0$. But starting
from a field of 17 Tesla it changes to $F_z=-2$. In the absence of
spin-orbit coupling [Fig.~\ref{coulomb}(a)] we still have several
crossing points between the levels with same total momentum.
Spin-orbit interaction [Fig.~\ref{coulomb}(b) and (c)] again mixes
those states and instead of level crossings we now have
`anticrossings'. Comparing with the single-hole states we notice
that the Coulomb interaction reduces the values of the level
repulsion gaps somewhat, but they are still orders of magnitude
larger than those for interacting electrons in a quantum dot
\cite{chakraborty}. Clearly, the Rashba effect has a much more
profound influence on hole quantum dots than that for electrons
which would manifest itself in a measurably quantity, such as
tunneling current in the dot \cite{Hong}.

\acknowledgments
The work was supported by the Canada Research
Chairs Program and the NSERC Discovery Grant.

\end{document}